\documentclass[]{jfm}

\usepackage[utf8]{inputenc}			
\usepackage[T1]{fontenc}			
\usepackage[british]{babel}	
\usepackage{amsmath}				
\usepackage{amssymb}				
\usepackage{bm}						
\usepackage{graphicx}				
\usepackage{xcolor} 				
\usepackage{comment} 				
\usepackage{array}					
\usepackage{tabularx}				
\usepackage[english=british]{csquotes} 

\usepackage{hyperref}		
\hypersetup{hidelinks} 		
\hypersetup{
    colorlinks = true,
    urlcolor   = blue,
    citecolor  = black,
}

\usepackage{ragged2e}
\usepackage{newtxtext}
\usepackage{newtxmath}
\usepackage{natbib}
\newcommand{\RomanNumeralCaps}[1]
\linenumbers

\renewcommand{\Pr}		{\mathrm{Pr}} 

\newcommand{\Ra}   		{\mathrm{Ra}}

\newcommand{\RaD}   	{\mathrm{Ra}_{\textrm{D}}}

\newcommand{\Nu}   		{\mathrm{Nu}}

\renewcommand{\Re}		{\mathrm{Re}} 

\newcommand{\dTN}   	{\Delta T_{\textrm{N}}}

\begin{document}

\title{Supergranule aggregation: a Prandtl number-independent feature of constant heat flux-driven convection flows}
\author{Philipp P. Vieweg\aff{1}
  \corresp{\email{philipp.vieweg@tu-ilmenau.de}}}
\affiliation{\aff{1}Institute of Thermodynamics and Fluid Mechanics, Technische Universit\"at Ilmenau, Postfach 100565, D-98684 Ilmenau, Germany}
\date{\today}

\maketitle

\begin{abstract}
Supergranule aggregation, i.e., the gradual aggregation of convection cells to horizontally extended networks of flow structures, is a unique feature of constant heat flux-driven turbulent convection. In the present study, we address the question if this mechanism of self-organisation of the flow is present for any fluid. 
Therefore, we analyse three-dimensional Rayleigh-Bénard convection at a fixed Rayleigh number $\Ra \approx 2.0 \times 10^{5}$ across $4$ orders of Prandtl numbers $\Pr \in \left[ 10^{-2}, 10^{2} \right]$ by means of direct numerical simulations in horizontally extended periodic domains with aspect ratio $\Gamma = 60$. 
Our study confirms the omnipresence of the mechanism of supergranule aggregation for the entire range of investigated fluids. Moreover, we analyse the effect of $\Pr$ on the global heat and momentum transport, and clarify the role of a potential stable stratification in the bulk of the fluid layer.
The ubiquity of the investigated mechanism of flow self-organisation underlines its relevance for pattern formation in geophysical and astrophysical convection flows, the latter of which are often driven by prescribed heat fluxes.
\end{abstract}

\section{Introduction}
\label{sec:Introduction}
Buoyancy, i.e., the interplay of gravity with mass density inhomogeneities that are typically caused by thermal heterogeneities, is, howsoever introduced, the essential mechanism that drives heat transport in many natural flows. 
Examples for such natural convection processes can be found on Earth throughout its layers from mantle convection \citep{Christensen1995} over deep ocean convection \citep{Maxworthy1994} up to convection in its atmosphere \citep{Atkinson1996}, eventually determining local and global aspects of weather and climate.

Natural thermal convection flows reveal often a hierarchy of different flow structures such as clusters of clouds over the warm ocean in the tropics of Earth \citep{Mapes1993}. The probably most prominent and thoroughly studied example of a hierarchy formation might be given by the solar convection zone in the outer $30 \%$ of the Sun \citep{Schumacher2020}. In this case, so-called \textit{granules} are superposed to larger flow structures termed \textit{supergranules}: although both of them are driven by the heat flux at the solar surface \citep{Schumacher2020, Rincon2018}, 
they offer very different lifetimes and horizontal extensions. Unfortunately, our understanding of such hierarchies' origins is still far from complete \citep{Hanson2020} and simpler set-ups of convection become necessary to improve it systematically. 

Rayleigh-Bénard convection represents the simplest conceivable set-up and, thus, the paradigm of naturally forced, thermally driven turbulence. Here, a fluid layer of thickness $H$ is confined between a heated horizontal plane at the bottom and a cooled one at the top -- because of the variation of density with temperature, the layer becomes unstable once subjected to gravity. As a result of intense research over the past decades, it is well-known that such convection systems organise themselves even in the fully turbulent regime into prominent \textit{long-living large-scale flow structures}. Although clearly distinguishable from the universal smaller-scale turbulence or fluctuations on significantly shorter time scales, the nature of these large-scale flow structures is complex and depends instead on various external factors such as the strength of the thermal driving, the working fluid or the presence of additional physical mechanisms \citep{Vieweg2023a}.

Interestingly, only very recent research identified \textit{thermal boundary conditions} as the key factor in determining the nature of these long-living large-scale flow structures given a horizontally extended domain. In a nutshell, either so-called \textit{turbulent superstructures} with characteristic horizontal extensions of $\Lambda \sim \textit{O} \left( H \right)$ form \citep{Pandey2018, Stevens2018, Krug2020, Vieweg2023}, or a so-called \textit{gradual supergranule aggregation} takes place that might result in a domain-sized flow structure with $\Lambda \gg \textit{O} \left( H \right)$ if not being interrupted by additional mechanisms such as rotation around the vertical axis \citep{Vieweg2021, Vieweg2022b}. Although the former establish whether the horizontal planes offer uniform temperatures (so-called Dirichlet conditions), the latter correspond to planes that prescribe a uniform vertical temperature gradient or, in other words, a spatially constant heat flux (Neumann conditions). Furthermore, the supergranules are superposed to significantly smaller (yet large-scale) granule-like flow structures, so a hierarchy of different horizontally extended flow structures may establish even in a simple turbulence configuration. This effect of thermal boundary conditions extends also to the Lagrangian material transport and the present coherent features in the flow \citep{Vieweg2021a, Vieweg2022, Vieweg2024}. Remarkably, these different self-organisations of the flows persist across the entire numerically accessible range of Rayleigh numbers $\Ra \lesssim 10^{8}$ (which quantify the strength of the thermal driving) \citep{Vieweg2021, Vieweg2022b}. Hence, the way how buoyancy effects are prescribed at the planes or boundaries seems to eventually determine the large-scale nature of the flows in between.

\begin{figure}
\centering
\includegraphics[scale = 1.0]{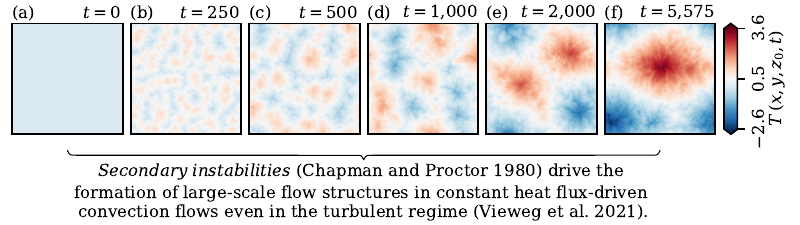}
\caption{\justifying{Gradual supergranule aggregation. 
Although secondary instabilities are essential for the transient growth of the supergranules, the final flow resembles a state described by the primary instability \citep{Hurle1967}. This time series visualises a flow at $\Pr = 10^{-2}$ (see table \ref{tab:simulation_parameters}) across the entire horizontal cross-section of aspect ratio $\Gamma = 60$ at $z_{0} = 1 - \delta_{T} /2$ with the thermal boundary-layer thickness $\delta_{T} = 1 / \left( 2 \thickspace \Nu \right)$.}}
\label{fig:gradual_supergranule_aggregation}
\end{figure}

Exceeding a critical value of thermal driving, the buoyancy-induced destabilisation leads to an onset of convection. Although this critical value depends on the thermal boundary conditions, the latter modify in particular the horizontal extension of the emerging flow structures. In more detail, this \textit{primary} instability leads to the emergence of convection rolls that exhibit predominantly one particular horizontal extension: depending on the mechanical boundary conditions, the corresponding critical wave numbers are $k_{\textrm{h, crit}} = \left[ 2.22, 3.13 \right]$ \citep{Rayleigh1916, Pellew1940} or $k_{\textrm{h, crit}} = 0$ \citep{Hurle1967} for applied uniform temperatures or vertical temperature gradients, respectively. This latter value is further supported by the \textit{secondary} instability slightly above the onset of convection, revealing that \enquote{each mode is unstable to one of longer wavelength than itself, so that any long box will eventually contain a single roll} \citep{Chapman1980}. 
In other words, any convection roll (of arbitrary size) is, at least slightly above the onset of convection, unstable to a more extended convection roll if buoyancy is introduced via a constant heat flux. Given that this result is obtained from a nonlinear evolution equation for the two-dimensional leading-order temperature perturbation, it is remarkable that a three-dimensional leading Lyapunov vector stability analysis discovered for a Prandtl number $\Pr = 1$ (which defines the working fluid) that the gradual supergranule aggregation is, even far beyond the onset of convection, driven by such secondary instabilities \citep{Vieweg2021, Chapman1980}, see also figure \ref{fig:gradual_supergranule_aggregation}. Once the numerically finite horizontal extent of the domain is reached, the \textit{final} statistically stationary state resembles essentially a finite-size relic of critical mode and thus shares similarities with the primary instability. 
Crucially, the latter is independent of the working fluid, whereas secondary and subsequent instabilities depend at least in the classical case of prescribed temperatures strongly on the working fluid \citep{Busse1978, Busse2003}. 
In the case of a prescribed heat flux, the authors studying secondary instabilities stated that their \enquote{results hold quite generally for all Prandtl numbers} \citep{Chapman1980a} 
but simultaneously \enquote{do not expect the theory to remain accurate for very small $\Pr$} \citep{Chapman1980}. 
As the final supergranule results from the preceding transient supergranule aggregation, clarifying this uncertainty becomes crucial especially due to the strongly varying Prandtl numbers in geophysical and astrophysical flows.

In the present work, we conduct direct numerical simulations across an extended range of fluids applicable to geophysical and astrophysical convection systems while prescribing constant vertical temperature gradients at the horizontal top and bottom planes. Providing extraordinarily long evolution times of up to the order of $\mathit{O} \left( 10^{4} \right)$ convective time units, we confirm that supergranule aggregation is an omnipresent feature independently of the working fluid. Despite its involved hierarchy of different large-scale flow structures, the global heat and momentum transport of the flows shares clear analogies with the complementary turbulent superstructures that manifest in the case of applied constant temperatures. 
Interestingly, the bulk stratification might manifest qualitatively differently depending on the working fluid. 

\section{Numerical method}
\label{sec:Numerical_method}

We consider the simplest conceivable scenario of convection based on the Oberbeck-Boussinesq approximation \citep{Oberbeck1879, Boussinesq1903} where the key idea is that the dependence of material parameters on \enquote{pressure is unimportant and that even the variation with temperature may be disregarded except in so far as it modifies the operation of gravity} \citep{Rayleigh1916}. 
As a consequence, the mass density $\rho$ becomes a linear function of only the temperature when it acts together with gravity but is constant or incompressible otherwise.

The three-dimensional equations of motion are solved by the spectral-element method Nek5000 \citep{Fischer1997,Scheel2013}. The equations are made dimensionless based on the layer height $H$ and the applied constant vertical temperature gradient $\beta$ at the plates, resulting in $\beta H$ as the characteristic temperature scale. Together with the free-fall inertial balance, the free-fall velocity $U_{\textrm{f}} = \sqrt{g \alpha \beta H^{2}}$ and free-fall time scale $\tau_{\textrm{f}} = 1 / \sqrt{\alpha g \beta}$ establish as characteristic units. Here, $\alpha$ is the volumetric thermal expansion coefficient at constant pressure and $g$ the acceleration due to gravity. 
This translates the equations eventually into
\begin{align}
\label{eq:CE}
\nabla \cdot \bm{u} &= 0 , \\
\label{eq:NSE}
\frac{\partial \bm{u}}{\partial t} + \left( \bm{u} \cdot \nabla \right) \bm{u} &= - \nabla p + \sqrt{\frac{\Pr}{\Ra}} \thickspace \nabla^{2} \bm{u} + T \bm{e}_{z} , \\
\label{eq:EE}
\frac{\partial T}{\partial t} + \left( \bm{u} \cdot \nabla \right) T &= \frac{1}{\sqrt{\Ra \Pr}} \thickspace \nabla^{2} T 
\end{align}
with $\bm{u}$, $T$ and $p$ representing the velocity, temperature and pressure field, respectively. The relative strength of the individual terms in these equations is controlled by the Rayleigh and Prandtl number, 
\begin{equation}
\label{eq:def_Rayleigh_number_Prandtl_number}
\Ra := \frac{\alpha g \beta H^{4}}{\nu \kappa} \qquad \textrm{and} \qquad \Pr := \frac{\nu}{\kappa} ,
\end{equation}
only. The quantities $\nu$ and $\kappa$ denote the viscosity and thermal diffusivity, respectively, and thus define the strength of molecular diffusion processes.

Independently of $\Ra$ and $\Pr$, equations \eqref{eq:CE} -- \eqref{eq:EE} are complemented by a three-dimensional domain with a square horizontal cross-section $A = \Gamma \times \Gamma$ and an aspect ratio $\Gamma := L / H = 60$ where $L$ is the horizontal periodic length of the domain. 
We apply at the top and bottom planes mechanical free-slip boundary conditions
\begin{equation}
\label{eq:def_free_slip_BC}
u_{z} \thickspace \left( z \in \left\{ 0, 1 \right\} \right) = 0 , 
\quad 
\frac{\partial u_{x, y}}{\partial z} \thickspace \left( z \in \left\{ 0, 1 \right\} \right) = 0, 
\end{equation}
as well as thermal constant heat flux boundary conditions
\begin{equation}
\label{eq:def_constant_flux_BC}
\frac{\partial T}{\partial z} \thickspace \left( z \in \left\{ 0, 1 \right\} \right) = - 1 .
\end{equation}

In spite of our interest in \textit{large}-scale flow structures, our direct numerical simulations resolve all dynamically relevant scales of the flows ranging from the domain size down to the dissipation scales based on a (refined) Grötzbach criterion \citep{Scheel2013}. These dissipation scales are given by the so-called Kolmogorov and Batchelor scale \citep{Kolmogorov1991, Batchelor1959, Sreenivasan2004},
\begin{equation}
\label{eq:def_Kolmogorov_and_Batchelor_scale}
\eta_{\textrm{K}} := \frac{\Pr^{3/8}}{\Ra^{3/8} \thickspace \varepsilon^{1/4}} 
\quad \textrm{and} \quad 
\eta_{\textrm{B}} := \frac{\eta_{\textrm{K}}}{\sqrt{\Pr}} ,
\end{equation}
for the velocity and scalar temperature field, respectively, where $\varepsilon := \left( 1 / 2 \right) \sqrt{\Pr / \Ra} \thickspace \big[ \left( \nabla \bm{u} \right) + \left( \nabla \bm{u} \right)^{T} \big]^{2}$ represents the kinetic energy dissipation rate. Note that whereas the Batchelor scale $\eta_{\textrm{B}} \leq \eta_{\textrm{K}}$ applies for $\Pr \geq 1$, the Corrsin scale $\eta_{\textrm{C}} := \eta_{\textrm{K}} / \Pr^{3/4}$ \citep{Corrsin1951} is here not of particular interest as it applies only at $\Pr \leq 1$ where $\eta_{\textrm{C}} \geq \eta_{\textrm{K}}$.

\section{Results}
\label{sec:Results}

In contrast to our previous work \citep{Vieweg2021}, we fix here the Rayleigh number $\Ra \approx 2.0 \times 10^{5}$ but vary instead the Prandtl number $\Pr \in \left[ 10^{-2}, 10^{2} \right]$ across $4$ orders of magnitude centred around $\Pr = 1$. The precise parameters are summarised for all our simulation runs in table \ref{tab:simulation_parameters}.

\newcommand{\hp}{\hphantom{1}}
\newcommand{\hpp}{\hphantom{.01}}
\begin{table}
\centering
\begin{tabular}{@{\hskip 0mm} r @{\hskip 3.2mm} c @{\hskip 3.2mm} r @{\hskip 3.2mm} r @{\hskip 3.2mm} r @{\hskip 3.2mm} r @{\hskip 3.2mm} r @{\hskip 3.2mm} r @{\hskip 3.2mm} r @{\hskip 3.2mm} c @{\hskip 3.2mm} c @{\hskip 0mm}}
\multicolumn{1}{c}{$\Pr \quad$} & \multicolumn{1}{c}{$N_{\textrm{e}} \quad$}	& \multicolumn{1}{c}{$N \thickspace$}	& \multicolumn{1}{c}{$t_{\textrm{r}} \left[ \tau_{\textrm{f}} \right]$}	& \multicolumn{1}{c}{$t_{\textrm{r}} \left[ \tau_{\nu} \right]$}	& \multicolumn{1}{c}{$t_{\textrm{r}} \left[ \tau_{\kappa} \right]$}	& \multicolumn{1}{c}{$\Lambda_{T} \thickspace \thickspace$} & \multicolumn{1}{c}{$\Nu \quad$}	& \multicolumn{1}{c}{$\Re \quad$} & \multicolumn{1}{c}{$\langle \eta_{\textrm{K}} \rangle_{V, t} \thickspace \thickspace \thickspace$} & \multicolumn{1}{c}{$\langle \eta_{\textrm{B}} \rangle_{V, t}$} \\ [3pt]
$0.01$		& $830^{2} \times 16$		& $13$	&  $ 5,575$		& $  1.2$    & $123.6$    &  $59.7$ &  $3.17$	&  $2063.0$	& $5.2 \times 10^{-3}$    & $5.2 \times 10^{-2}$ \\
$0.1 \hp$	& $400^{2} \times \hp 8$	& $9$	&  $ 4,250$		& $  3.0$    & $ 29.8$    &  $59.7$ &  $4.94$	&  $433.0$  & $1.6 \times 10^{-2}$    & $5.0 \times 10^{-2}$ \\
$1 \hpp$ 	& $200^{2} \times \hp 4$	& $11$	&  $ 6,500$     & $ 14.4$    & $ 14.4$    &  $59.7$ &  $6.74$	&  $81.4$	& $4.9 \times 10^{-2}$    & $4.9 \times 10^{-2}$ \\
$7 \hpp$	& $200^{2} \times \hp 4$	& $7$	&  $ 4,000$		& $ 23.5$    & $  3.4$    &  $59.8$ &  $7.21$	&  $16.2$	& $1.3 \times 10^{-1}$    & $4.9 \times 10^{-2}$ \\
$10 \hpp$	& $200^{2} \times \hp 4$	& $7$	&  $ 6,000$		& $ 42.1$    & $  4.2$    &  $59.8$ &  $7.13$	&  $11.7$	& $1.5 \times 10^{-1}$    & $4.9 \times 10^{-2}$ \\
$100 \hpp$	& $200^{2} \times \hp 4$	& $7$	&  $14,000$		& $310.3$    & $  3.1$    &  $59.8$ &  $7.02$	&  $1.1$	& $4.9 \times 10^{-1}$    & $4.9 \times 10^{-2}$ \\
\end{tabular}
\caption{\justifying{Simulation parameters
of the direct numerical simulations at different Prandtl numbers $\Pr$ -- the Rayleigh number $\Ra = 203,576$, aspect ratio $\Gamma = 60$, and free-slip as well as constant heat flux boundary conditions are applied for all runs. 
The table contains further the total number of spectral elements $N_{\textrm{e}}$ in the simulation domain, the polynomial order $N$ on each spectral element, the total runtime of the simulation $t_{\textrm{r}}$ in units of the corresponding free-fall times $\tau_{\textrm{f}}$, a subsequent translation of these runtimes into vertical diffusion times $\tau_{\nu, \kappa}$, as well as the resulting integral length scale $\Lambda_{T}$ of the temperature field at midplane, Nusselt number $\Nu$, Reynolds number $\Re$, and the mean Kolmogorov and Batchelor scale, $\langle \eta_{\textrm{K}} \rangle_{V, t}$ and $\langle \eta_{\textrm{B}} \rangle_{V, t}$, respectively. 
$\Lambda_{T}$, $\Nu$, $\Re$, $\langle \eta_{\textrm{K}} \rangle_{V, t}$, and $\langle \eta_{\textrm{B}} \rangle_{V, t}$ are obtained from the last $500 \tau_{\textrm{f}}$ ($5 \tau_{\textrm{f}}$ for $\Pr = 10^{-2}$, $1000 \tau_{\textrm{f}}$ for $\Pr = 10^{2}$) of each simulation run.
}}
\label{tab:simulation_parameters}
\end{table}
\let\hp\undefined
\let\hpp\undefined

\subsection{Ubiquitous gradual supergranule aggregation}
\label{subsec:Ubiquitous_gradual_supergranule_aggregation}

Initialised with its fluid at rest possessing a randomly perturbed linear diffusive equilibrium profile, i.e., $\bm{u} \left( t = 0 \right) = 0$ and $T \left( t = 0 \right) = T_{\textrm{lin}} + \Psi$ together with $T_{\textrm{lin}} := 1 - z$ and $0 \leq \Psi \left( \bm{x} \right) \leq 10^{-3}$ \citep{Vieweg2023a, Vieweg2024}, every simulation is run as long as necessary to indicate a stationarity of the large-scale flow structure formation. This can be captured, for instance, by (i) the thermal variance $\Theta_{\textrm{rms}}$ with the temperature deviation $\Theta = T - T_{\textrm{lin}}$ or (ii) the integral length scale \citep{Parodi2004} 
of the temperature field 
\begin{equation}
\label{eq:def_integral_length_scale}
\Lambda_{T} \left( z_{0}, t \right) := 2 \pi \frac{\int_{k_{\textrm{h}}} \left[ E_{TT} \left( k_{\textrm{h}}, z_{0}, t \right) / k_{\textrm{h}} \right] dk_{\textrm{h}}}{\int_{k_{\textrm{h}}} E_{TT} \left( k_{\textrm{h}}, z_{0}, t \right) dk_{\textrm{h}}} 
\end{equation}
based on the azimuthally averaged Fourier energy spectrum at midplane, $E_{TT} \left( k_{\textrm{h}}, z_{0} = 0.5, t \right)$, as shown in \citep{Vieweg2022b}. Note here that neither the Reynolds nor the Nusselt number (see below in equations \eqref{eq:def_Reynolds_number} and \eqref{eq:def_Nusselt_number}, respectively) reflect the transient large-scale structure formation properly \citep{Vieweg2021, Vieweg2022b}. Running our simulations reveals two particularly interesting results.

\begin{figure}
\centering
\includegraphics[scale = 1.0]{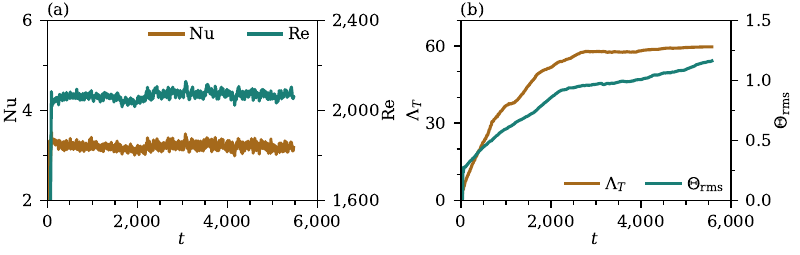}
\caption{\justifying{Signs of the transient supergranule aggregation. 
(a) While neither $\Nu$ nor $\Re$ are affected significantly, (b) $\Lambda_{T}$ and $\Theta_{\textrm{rms}}$ do indicate the transient supergranule aggregation. The data correspond to $\Pr = 10^{-2}$, see also figure \ref{fig:gradual_supergranule_aggregation}.}}
\label{fig:evolution_Nu_Re_intls_rms_Pr001}
\end{figure}

First, the gradual supergranule aggregation, first reported in \citep{Vieweg2021}, sets in even beyond $\Pr = 1$ at all accessible Prandtl numbers as both $\Theta_{\textrm{rms}}$ and $\Lambda_{T}$ increase over time, see also figure \ref{fig:evolution_Nu_Re_intls_rms_Pr001}. Yet, the varying diffusivities affect the pace of the dynamics and thus the necessary simulation runtime $t_{\textrm{r}}$ to reach a statistically stationary large-scale pattern size, see table \ref{tab:simulation_parameters}. Although $t_{\textrm{r}}$ is by far largest for the upper investigated limit of $\Pr$, we find a similar trend in the opposing lower limit. This observation confirms that the efficiency of the aggregation process depends on the interplay of the velocity and temperature field, being in line with our previous results \citep{Vieweg2022b} which trace the (thermal) supergranule aggregation basically back to an \textit{advective transfer of thermal variance}. Consequently, these runtimes do not support any relation to the diffusive time scales $\tau_{\nu} = H^{2} / \nu \equiv \sqrt{\Ra / \Pr} \thickspace \tau_{\textrm{f}}$ and $\tau_{\kappa} = H^{2} / \kappa \equiv \sqrt{\Ra \Pr} \thickspace \tau_{\textrm{f}}$ as contrasted in table \ref{tab:simulation_parameters}. Interestingly, while more simulations are required to draw firm conclusions on the interplay of diffusion processes concerning the pace of the aggregation process, the increase of necessary runtime is clearly larger in the direction $\Pr \rightarrow \infty$.

\begin{figure}
\centering
\includegraphics[scale = 1.0]{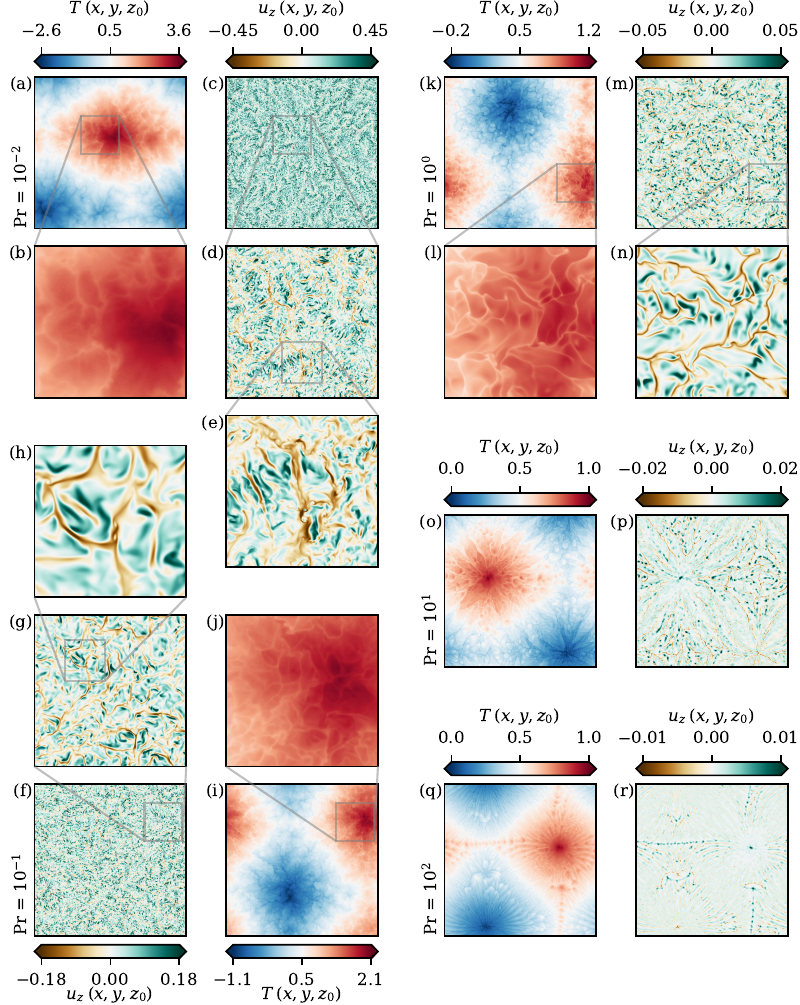}
\caption{\justifying{Supergranulation across four orders of Prandtl numbers. 
Although the velocity field exhibits successively smaller features for decreasing Prandtl numbers $\Pr$, the supergranule aggregation can still easily be observed in the temperature field.
Panels (a, c, f, i, k, m, o, p, q, r) visualise the entire cross-section at $z_{0} = 1 - \delta_{T} /2$. To highlight the vast scale-separation between the temperature and (vertical) velocity field for small $\Pr$, panels (b, d, g, j, l, m) show enlarged regions of interest of size $15 \times 15$. Panels (e, h) underline this fact by additional magnifications of regions of size $4 \times 4$.}}
\label{fig:supergranulation_at_different_Pr}
\end{figure}

Second, this process ceases, independently of $\Pr$, only once the horizontal domain size is reached, implying that thermal variance has significantly aggregated on the scale of the horizontal domain size. Consequently, the integral length scale $\Lambda_{T}$ converges in any simulation run towards $\Gamma$ as indicated by table \ref{tab:simulation_parameters} and figure \ref{fig:evolution_Nu_Re_intls_rms_Pr001} (b). 
Figure \ref{fig:supergranulation_at_different_Pr} visualises the temperature and vertical velocity field in horizontal planes within the upper thermal boundary layer for these final states of the flows. In particular, panels (a, i, k, o, q) depict the temperature fields across the entire horizontal cross-sections of the domains, whereas panels (c, f, m, p, r) exemplary contrast them to the velocity field with respect to its vertical component. 
The circumstance that the supergranules grow in every run without any upper physical limit confirms that the secondary instability mechanism \citep{Chapman1980, Chapman1980a} rules the formation of long-living large-scale flow structures even far beyond the onset of convection independently of $\Pr$.

\begin{figure}
\centering
\includegraphics[scale = 1.0]{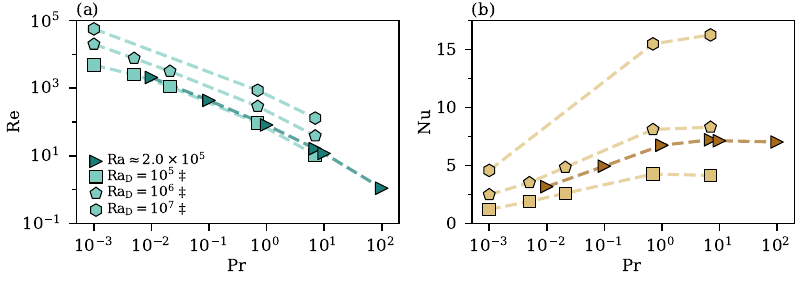}
\caption{\justifying{Global momentum and heat transport for different fluids. 
(a) While the global momentum transport increases with decreasing $\Pr$, (b) the convective heat transport loses importance (relative to purely diffusive heat transport) only for $\Pr \lesssim 1$. The \textit{dark} markers correspond to supergranule data from the late state of this study's flows as described by table \ref{tab:simulation_parameters}. In contrast, the \textit{bright} markers represent turbulent superstructure data (i.e., different thermal boundary conditions) as outlined in the discussion.
In a nutshell, the present study differs from $\ddagger$\citep{Pandey2022} as follows: 
thermal Neumann boundary conditions vs. Dirichlet conditions,
horizontally periodic domain of $\Gamma = 60$ vs. closed box of $\Gamma = 25$,
mechanical free-slip boundary conditions at the top and bottom planes vs. no-slip conditions.
Note that the series at $\Ra \approx 2.0 \times 10^{5}$ and $\RaD = 10^{5}$ can be related \citep{Vieweg2023a}.
}}
\label{fig:scaling_and_comparison_global_transport}
\end{figure}

Albeit the gradual supergranule aggregation seems to be a ubiquitous feature across all covered fluids, the variation of the Prandtl number still modifies other aspects of the flow. While they display well-ordered stems of localised up- and down-flow regions for large $\Pr$, they become increasingly disordered for increasingly smaller $\Pr$ due to the reduced importance of molecular friction. Consequently, the ranges of observable scales or details diverge when comparing the temperature and vertical velocity field: this is highlighted in figure \ref{fig:supergranulation_at_different_Pr} by magnifications of fractions of the flows. In the case of $\Pr = 1$, both fields offer an equivalent richness of details which is shown in panels (l, n). This changes once the Prandtl number moves off unity and the diffusivities of momentum and the scalar temperature or the mean Kolmogorov \citep{Kolmogorov1991} and Batchelor \citep{Batchelor1959} scales differ. On the one hand, the temperature field becomes successively diffuse or imprecise for increasingly smaller $\Pr$, compare thereto panels (b, j, l). On the other hand, the velocity field becomes simultaneously successively more chaotic as directly contrasted in panels (d, g, n). The tremendous scale separation between the two fields is ultimately highlighted by further magnifications of even smaller regions in panels (e, h), underlining the vast complexity of low-$\Pr$ thermal convection flows. Table \ref{tab:simulation_parameters} quantifies this visual scale separation by including the mean Kolmogorov and Batchelor scale for each simulation.

The increasing local disparity of the temperature and velocity field due to the different time scales of the underlying diffusion processes indicates that the impact of a variation of $\Pr$ on the global transport of momentum and heat should be investigated in the following.

\subsection{Global transport properties and the role of stratification}
\label{subsec:Global_transport_properties_and_the_role_of_stratification}

An alternative perspective on the response of the dynamical system on its buoyancy-induced forcing is provided by its global momentum and heat transport as can be measured by the Reynolds and Nusselt number, respectively. While the former is given by
\begin{equation}
\label{eq:def_Reynolds_number}
\Re \thickspace (t) := \sqrt{\frac{\Ra}{\Pr}} \thickspace u_{\textrm{rms}} \qquad \textrm{with } u_{\textrm{rms}} := \sqrt{ \langle \bm{u}^{2} \rangle_{V} } ,
\end{equation}
the latter quantifies the strength of convective heat transport, by comparing the total heat transport across the fluid layer to a state of pure heat conduction, and results (in the present case of an applied constant heat flux) in \citep{Otero2002}
\begin{align}
\label{eq:def_Nusselt_number}
\Nu \thickspace (t) = \frac{1}{\dTN} \qquad \textrm{with } \dTN := \langle T(z = 0) - T(z = 1) \rangle_{A} \leq 1
\end{align}
where $\dTN$ is the dynamically manifesting mean temperature drop across the fluid layer. 

Figure \ref{fig:scaling_and_comparison_global_transport} visualises via dark markers the dependence of these global transport measures on the Prandtl number for the final flow states, see again figure \ref{fig:supergranulation_at_different_Pr}. 
On the one hand, the Reynolds number can be found to increase steadily when the Prandtl number is decreased. This is in accordance with the vanishing role of viscous diffusion, allowing for higher velocities and leading to successively more inertial flows. As this holds for the entire range of covered Prandtl numbers, it implies that the flow laminarises for $\Pr \gg 1$.
On the other hand, the Nusselt number shows a more complex behaviour. For decreasing Prandtl numbers in the range $\Pr \lesssim 1$, thermal diffusion gains relevance as the disorder in the flow intensifies (see $\Re$). In contrast, $\Nu$ stagnates for $\Pr \gtrsim 1$: this effect might be induced by the full nesting of the thermal boundary layer into the viscous one \citep{Chilla2012} 
(the latter of which might be estimated to be $\delta_{u} \sim \Pr \thickspace \delta_{T}$ based on diffusion arguments), so buoyancy effects are suppressed or protracted by viscous diffusion and thermal plumes detach less frequently. 

Thermal plume detachments are fundamentally caused by the applied (inverse or) \textit{unstable density stratification} introduced at the heated bottom and cooled top plane. These ascending and descending plumes leave consequently the boundary layers and travel, driven by buoyancy, into or even through the bulk, leading to turbulent mixing once the flow is sufficiently inertial. Remarkably, our previous study \citep{Vieweg2021} observed a slightly \textit{stable density stratification} for any constant heat flux-driven convection flow in the bulk region independently of $\Ra$ given $\Pr = 1$. In other words, the flow structures established a density stratification that was counter-directed to the applied one. Although the strength of this stratification decreased with increasing $\Ra$, it remained stable for all accessible $\Ra \lesssim 10^{8}$.
In the following, we address the question of whether such a stable stratification is a unique feature of every flow that exhibits the effect of supergranule aggregation.

\begin{figure}
\centering
\includegraphics[scale = 1.0]{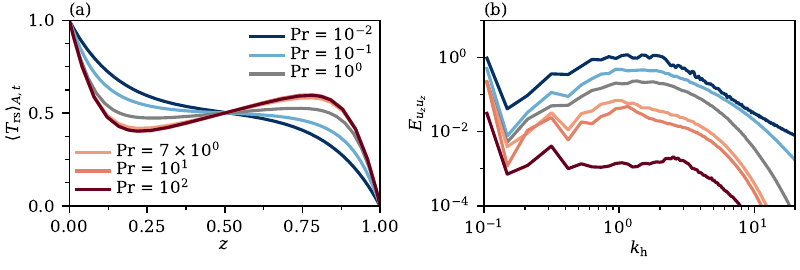}
\caption{\justifying{Stratification and the dominance of supergranular flow structures in different fluids. 
(a) A stable stratification in the bulk is no omnipresent result of (or necessity for) the emergence of the supergranule but rather related to plume detachments. (b) Simultaneously, the supergranules at $k_{\textrm{h, min}} = 2 \pi / \Gamma \approx 0.1$ become, evaluating here the midplane, weaker relative to smaller-scale structures (such as granules) for decreasing $\Pr$. 
The temperature fields are re-scaled as described in the text, and both data correspond to the late state of the flows as described in the caption of Table \ref{tab:simulation_parameters}.}}
\label{fig:temperature_profiles_spectral_peaks}
\end{figure}

Therefore, we contrast the temperature profiles of all present runs in figure \ref{fig:temperature_profiles_spectral_peaks} (a). Note that the temperature fields are re-scaled here via $T_{\textrm{rs}} = \left( T - \langle T \rangle_{V} \right) / \dTN + \langle T \rangle_{V}$ (which does not affect the stratification properties) to allow for a direct comparison. Unlike in our previous study, we find here stably as well as unstably stratified bulks despite the presence of supergranules for any $\Pr$. While it is stable for $\Pr \geq 1$ and converges for $\Pr \gtrsim 7$, it is increasingly unstable for successively smaller Prandtl numbers $\Pr < 1$. Interestingly, these trends coincide with the above findings regarding the scaling of $\Nu \left( \Pr \right)$, suggesting a relation of the bulk stratification with plume detachments. Hence, a stable stratification in the bulk is no omnipresent result of the emergence of supergranular flow structures, while the potentially forming local peaks in the temperature profile can be seen as the consequence of a competition between the protracted overshooting thermal plumes and the opposite boundary layers close to the top and bottom plane.

\section{Discussion and perspective}

Introducing buoyancy in a simple Rayleigh-Bénard convection configuration via a constant heat flux at the top and bottom planes leads without any additional physics to the emergence of a hierarchy of different long-living large-scale flow structures \citep{Vieweg2021, Vieweg2023a}. While this hierarchy consists of so-called granules and supergranules as separate stages, the latter are driven by secondary instabilities and might grow until the horizontal domain size is reached (see again section \ref{sec:Introduction} and Figure \ref{fig:gradual_supergranule_aggregation}). 
The present study raises the question if mechanisms similar to this secondary instability in constant heat flux-driven Rayleigh-Bénard convection \citep{Chapman1980a, Chapman1980} 
act beyond $\Pr = 1$ \citep{Vieweg2021} independently of the working fluid and, in particular, even down to very small Prandtl numbers such as found in the solar convection zone \citep{Schumacher2020, Rincon2018}. We therefore conducted a series of simulations across four orders of $\Pr$ given a fixed thermal driving in a horizontally extremely extended periodic domain of $\Gamma = 60$. We confirmed the presence of the gradual supergranule aggregation as a particular mechanism of self-organisation of long-living large-scale flow structures in naturally forced convection flows independently of the working fluid. Our observations thus suggest that these secondary instabilities dominate any basic heat flux-driven convection flow, leading to a robust hierarchy of different large-scale flow structures. 

This omnipresent appearance is in accordance with the accessibility of large-scale $k_{z} = 0$ spectral modes in the temperature field for this particular thermal boundary condition. Note that these modes are not accessible in the classical case of applied constant temperatures \citep{Vieweg2022b}. As the present configuration corresponds to a ratio of thermal diffusivities $\kappa_{\textrm{f}} / \kappa_{\textrm{s}} \rightarrow \infty$ between the fluid and the above or below solid, this mechanism can be seen as the result of a relaxation of thermal perturbations that happens much quicker in the fluid compared to in the solid plates \citep{Hurle1967, Vieweg2023}. Moreover, the strength of buoyancy effects is in the heat flux-driven scenario not limited by prescribed temperatures at the boundaries but instead via only the less-restrictive mixing of fluid in between. Hence, these arguments allow and demand eventually the formation of horizontally extended flow structures that might even span across the entire domain to advectively transfer the thermal variance.

Given the fact that the variation of the working fluid affects the relative strength of thermal diffusion as described by equation \eqref{eq:def_Rayleigh_number_Prandtl_number}, one might expect a \textit{de}crease of thermal variance for decreasing $\Pr$ due to an increase of $\kappa$. However, it turns out that the thermal variance \textit{in}creases -- this is also indicated by the colour scales in figure \ref{fig:supergranulation_at_different_Pr}. This observation can be explained as follows: smaller $\Pr$ result in larger $\Re$ and thus in an increased \textit{local} mixing (with $\kappa$ acting also locally). As the flow is increasingly disordered, the large-scale supergranule becomes less dominant compared with smaller-scale velocity structures which is confirmed by the spectral analysis captured in figure \ref{fig:temperature_profiles_spectral_peaks} (b). Consequently, the horizontal mixing on large scales (see also the previous paragraph) becomes successively less effective for smaller $\Pr$, leading eventually to an increased thermal variance in the horizontally extended domain. 
As becomes clear when contrasting the present results with \citep{Vieweg2021}, (i) the vanishing stable stratification 
is not an effect of the increased Reynolds number and (ii) the relative heat transport of supergranules compared with smaller-scale structures (such as granules) loses similarly importance when increasing $\Re$ via $\Ra$ given a fixed $\Pr$.


Interestingly, despite the fundamentally different long-living large-scale flow structures between the cases of applied constant temperatures and vertical temperature gradients (see again section \ref{sec:Introduction}), their response on a variation of the working fluid shares clear analogies: compare therefore the bright and dark markers in figure \ref{fig:scaling_and_comparison_global_transport}, respectively. 
Note here that while the Rayleigh number $\RaD := \alpha g \Delta T H^{3} / \left( \nu \kappa \right)$ in case of an applied constant temperature difference $\Delta T$, this is related via $\RaD \equiv \Ra / \Nu$ \citep{Otero2002, Vieweg2021, Foroozani2021} to equation \eqref{eq:def_Rayleigh_number_Prandtl_number}. In particular, this allows to relate the present $\Pr = 1$ run to the corresponding no-slip and $\RaD = 10^{5}$ one from \citep{Pandey2022} as described in \citep{Vieweg2023a}. Thus, the particular kind of thermal boundary condition seems not to be of great significance when it comes to qualitative trends of the classical global measures of heat and momentum transport with respect to $\Pr$. In other words, different large-scale flow structures respond qualitatively similarly on a variation of the working fluid if judged via global measures of heat or momentum transport. 
Moreover, this underlines that diffusion processes are primarily \textit{locally} important and do not rule the large-scale pattern formation. 

The omnipresence of supergranule aggregation across all accessible Rayleigh and Prandtl numbers highlights the importance of an understanding of secondary (and subsequent) instabilities \citep{Chapman1980a, Chapman1980} slightly above the onset of convection. It is intriguing that such mechanisms survive even into the fully turbulent states of the flows \citep{Vieweg2021} where patterns are typically highly susceptible to the influence of instabilities and defects on each other \citep{Busse1978, Busse2003}. 
Moreover, additional physical mechanisms are required to stop the gradual supergranule aggregation before reaching the numerically finite domain size. Weak rotation around the vertical axis has turned out to effectively interrupt this process in the turbulent regime \citep{Vieweg2022b} while also the primary instability changes qualitatively with $k_{\textrm{h, crit}} > 0$ once rotation is sufficiently strong \citep{Dowling1988, Takehiro2002}. Interestingly, the ratio of thermal diffusivities $\kappa_{\textrm{f}} / \kappa_{\textrm{s}}$ seems to promise similar effects \citep{Hurle1967}. This is of particular importance to better resemble the motivating geophysical and astrophysical flows and will be addressed in future studies.

\backsection[Acknowledgements]{
The author thanks Prof. Jörg Schumacher for valuable comments on the early manuscript.
}

\backsection[Funding]
{
The author is supported by the Deutsche Forschungsgemeinschaft within the Priority Programme DFG-SPP 1881 \enquote{Turbulent Superstructures} as well as grant no. 1410/31-1. 
He gratefully acknowledges the Gauss Centre for Supercomuting e.V. (\href{www.gauss-centre.eu}{\texttt{www.gauss-centre.eu}}) for funding this work by providing computing resources on the GCS supercomputer SuperMUC-NG at Leibnitz Supercomputing Centre within project pn68ni and through the John von Neumann Institute for Computing (NIC) on the GCS supercomputer JUWELS at Jülich Supercomputing Center (JSC) within projects mesoc and nonbou.
Additionally, he acknowledges the computing centre of the Technische Universität Ilmenau for providing access to, as well as computing and storage resources on its compute cluster MaPaCC4.
}

\backsection[Declaration of interests]{
The author reports no conflict of interest.
}

\backsection[Author ORCIDs]{
P. P. Vieweg, https://orcid.org/0000-0001-7628-9902
}

\bibliographystyle{jfm}

\begin{thebibliography}{99}

\expandafter\ifx\csname natexlab\endcsname\relax
\def\natexlab#1{#1}\fi
\expandafter\ifx\csname selectlanguage\endcsname\relax
\def\selectlanguage#1{\relax}\fi

\bibitem[Atkinson and Wu Zhang (1996)]{Atkinson1996}
\textsc{Atkinson, B. W. and Wu Zhang, J.}
1996
Mesoscale Shallow Convection in the Atmosphere, 
\textit{Rev. Geophys.} \textbf{34}, 403--431.

\bibitem[Batchelor (1959)]{Batchelor1959}
\textsc{Batchelor, G. K.}
1959
Small-Scale Variation of Convected Quantities like Temperature in Turbulent Fluid Part 1. General Discussion and the Case of Small Conductivity, 
\textit{J. Fluid Mech.} \textbf{5}, 113.

\bibitem[Boussinesq (1903)]{Boussinesq1903}
\textsc{Boussinesq, J. V.}
1903
\textit{Théorie Analytique de La Chaleur}, 
Vol. 2 (Gauthier-Villars, Paris, France).

\bibitem[Busse (1978)]{Busse1978}
\textsc{Busse, F. H.}
1978
Non-Linear Properties of Thermal Convection,
\textit{Rep. Prog. Phys.} \textbf{41}, 1929--1967.

\bibitem[Busse (2003)]{Busse2003}
\textsc{Busse, F. H.}
2003
The Sequence-of-Bifurcations Approach towards Understanding Turbulent Fluid Flow
\textit{Surv. Geophys.} \textbf{24}, 269--288.

\bibitem[Chapman and Proctor (1980)]{Chapman1980}
\textsc{Chapman, C. J. and Proctor, M. R. E.}
1980
Nonlinear Rayleigh–Bénard Convection between Poorly Conducting Boundaries, 
\textit{J. Fluid Mech.} \textbf{101}, 759--782.

\bibitem[Chapman et al. (1980)]{Chapman1980a}
\textsc{Chapman, C. J., Childress, S. and Proctor, M. R. E.}
1980
Long Wavelength Thermal Convection between Non-Conducting Boundaries, 
\textit{Earth Planet. Sc. Lett.} \textbf{51}, 362--369.

\bibitem[Chilla and Schumacher (2012)]{Chilla2012}
\textsc{Chillà, F. and Schumacher, J.}
2012
New Perspectives in Turbulent Rayleigh-Bénard Convection, 
\textit{Eur. Phys. J. E} \textbf{35}, 58.

\bibitem[Christensen (1995)]{Christensen1995}
\textsc{Christensen, U.}
1995
Effects of Phase Transitions on Mantle Convection,
\textit{Annu. Rev. Earth Planet. Sci.} \textbf{23}, 65--87.

\bibitem[Corrsin (1951)]{Corrsin1951}
\textsc{Corrsin, S.}
1951
On the {{Spectrum}} of {{Isotropic Temperature Fluctuations}} in an {{Isotropic Turbulence}},
\textit{J. Appl. Phys.} \textbf{22}, 469--473.

\bibitem[Dowling (1988)]{Dowling1988}
\textsc{Dowling, T. E.}
1988
\textit{Rotating {{Rayleigh-Bénard Convection}} with {{Fixed Flux Boundaries}}} in \textit{1988 {{Summer Study Program}}
in {{Geophysical Fluid Dynamics}}: {{The Influence}} of {{Convection}} on {{Large-Scale Circulations}}} (Woods Hole Oceanographic Institution, Massachusetts), 230--247.

\bibitem[Fischer (1997)]{Fischer1997}
\textsc{Fischer, P. F.}
1997
An Overlapping Schwarz Method for Spectral Element Solution of the Incompressible Navier–Stokes Equations, 
\textit{J. Comput. Phys.} \textbf{133}, 84--101.

\bibitem[Foroozani et al. (2021)]{Foroozani2021}
\textsc{Foroozani, N., Krasnov, D. and Schumacher, J.}
2021
Turbulent Convection for Different Thermal Boundary Conditions at the Plates, 
\textit{J. Fluid Mech.} \textbf{907}, A27.

\bibitem[Hanson et al. (2020)]{Hanson2020}
\textsc{Hanson, C. S., Duvall, T. L., Birch, A. C., Gizon, L. and Sreenivasan, K. R.}
2020
Solar East-West Flow Correlations That Persist for Months at Low Latitudes Are Dominated by Active Region Inflows,
\textit{A\&A} \textbf{644}, A103.

\bibitem[Hurle et al. (1967)]{Hurle1967}
\textsc{Hurle, D. T. J., Jakeman, R. and Pike, E. R.}
1967
On the Solution of the Bénard Problem with Boundaries of Finite Conductivity,
\textit{Proc. R. Soc. Lond. A} \textbf{296}, 469--475.

\bibitem[Käufer et al. (2023)]{Vieweg2023}
\textsc{Käufer, T., Vieweg, P. P., Schumacher, J. and Cierpka, C.}
2023
Thermal Boundary Condition Studies in Large Aspect Ratio {{Rayleigh}}–{{Bénard}} Convection,
\textit{Eur. J. Mech. B-Fluids} \textbf{101}, 283--293.

\bibitem[Kolmogorov (1991)]{Kolmogorov1991}
\textsc{Kolmogorov, A. N.}
1991
The Local Structure of Turbulence in Incompressible Viscous Fluid for Very Large Reynolds Numbers,
\textit{Proc. Math. Phys. Sci.} \textbf{434}, 9--13.

\bibitem[Krug et al. (2020)]{Krug2020}
\textsc{Krug, D., Lohse, D. and Stevens, R. J. A. M.}
2020
Coherence of Temperature and Velocity Superstructures in Turbulent Rayleigh–Bénard Flow,
\textit{J. Fluid Mech.} \textbf{887}, A2.

\bibitem[Mapes and Houze (1993)]{Mapes1993}
\textsc{Mapes, B. E. and Houze, R. A.}
1993
Cloud Clusters and Superclusters over the Oceanic Warm Pool,
\textit{Mon. Wea. Rev.} \textbf{121}, 1398--1416.

\bibitem[Maxworthy and Narimousa (1994)]{Maxworthy1994}
\textsc{Maxworthy, T. and Narimousa, S.}
1994
Unsteady, Turbulent Convection into a Homogeneous, Rotating Fluid, with Oceanographic Applications,
\textit{J. Phys. Oceanogr.} \textbf{24}, 865--887.

\bibitem[Oberbeck (1879)]{Oberbeck1879}
\textsc{Oberbeck, A.}
1879
Ueber die Wärmeleitung der Flüssigkeiten bei Berücksichtigung der Strömungen infolge von Temperaturdifferenzen, 
\textit{Ann. Phys. Chem.} \textbf{243}, 271--292.

\bibitem[Otero et al. (2002)]{Otero2002}
\textsc{Otero, J., Wittenberg, R. W., Worthing, R. A., Doering, C. R.}
2002
Bounds on {{Rayleigh}}–{{Bénard}} Convection with an Imposed Heat Flux, 
\textit{J. Fluid Mech.} \textbf{473}, 191--199.

\bibitem[Pandey et al. (2018)]{Pandey2018}
\textsc{Pandey, A., Scheel, J. D. and Schumacher, J.}
2018
Turbulent Superstructures in Rayleigh-Bénard Convection,
\textit{Nat. Commun.} \textit{9}, 2118.

\bibitem[Pandey et al. (2022)]{Pandey2022}
\textsc{Pandey, A., Krasnov, D., Sreenivasan, K. R. and Schumacher, J.}
2022
Convective Mesoscale Turbulence at Very Low Prandtl Numbers, 
\textit{J. Fluid Mech.} \textbf{948}, A23.

\bibitem[Parodi et al. (2004)]{Parodi2004}
\textsc{Parodi, A., von Hardenberg, J., Passoni, G., Provenzale, A. and Spiegel, E. A.}
2004
Clustering of Plumes in Turbulent Convection, 
\textit{Phys. Rev. Lett.} \textbf{92}, 194503.

\bibitem[Pellew and Southwell (1940)]{Pellew1940}
\textsc{Pellew, A. and Southwell, R. V.}
1940
On Maintained Convective Motion in a Fluid Heated from Below, 
\textit{Proc. R. Soc. Lond. A} \textbf{176}, 312--343.

\bibitem[Rayleigh (1916)]{Rayleigh1916}
\textsc{Rayleigh, L.}
1916
On Convection Currents in a Horizontal Layer of Fluid, When the Higher Temperature Is on the under Side,
\textit{The London, Edinburgh, and Dublin Philosophical Magazine and Journal of Science} \textbf{32}, 529--546.

\bibitem[Rincon and Rieutord (2018)]{Rincon2018}
\textsc{Rincon, F. and Rieutord, M.}
2018
The Sun’s Supergranulation, 
\textit{Living Rev. Sol. Phys.} \textbf{15}, 6.

\bibitem[Scheel et al. (2013)]{Scheel2013}
\textsc{Scheel, J. D., Emran, M. S. and Schumacher, J.}
2013
Resolving the Fine-Scale Structure in Turbulent Rayleigh–Bénard Convection, 
\textit{New J. Phys.} \textbf{15}, 113063.

\bibitem[Schneide et al. (2022)]{Vieweg2022}
\textsc{Schneide, C., Vieweg, P. P., Schumacher, J. and Padberg-Gehle, K.}
2022
Evolutionary Clustering of {{Lagrangian}} Trajectories in Turbulent {{Rayleigh}}–{{Bénard}} Convection Flows,
\textit{Chaos} \textbf{32}, 013123.

\bibitem[Schumacher and Sreenivasan (2020)]{Schumacher2020}
\textsc{Schumacher, J. and Sreenivasan, K. R.}
2020
Colloquium: Unusual Dynamics of Convection in the Sun,
\textit{Rev. Mod. Phys.} \textbf{92}, 041001.

\bibitem[Sreenivasan (2004)]{Sreenivasan2004}
\textsc{Sreenivasan, K. R.}
2004
Possible {{Effects}} of {{Small-Scale Intermittency}} in {{Turbulent Reacting Flows}},
\textit{Flow Turbul. Combust.} \textbf{72}, 115--131.

\bibitem[Stevens et al. (2018)]{Stevens2018}
\textsc{Stevens, R. J. A. M., Blass, A., Zhu, X., Verzicco, R. and Lohse, D.}
2018
Turbulent Thermal Superstructures in Rayleigh-Bénard Convection, 
\textit{Phys. Rev. Fluids} \textbf{3}, 041501.

\bibitem[Takehiro et al. (2002)]{Takehiro2002}
\textsc{Takehiro, S.-I., Ishiwatari, M., Nakajima, K. and Hayashi, Y.-Y.}
2002
Linear {{Stability}} of {{Thermal Convection}} in {{Rotating Systems}} with {{Fixed Heat Flux Boundaries}}, 
\textit{Geophys. Astro. Fluid} \textbf{96}, 439--459.

\bibitem[Vieweg et al. (2021)]{Vieweg2021}
\textsc{Vieweg, P. P., Scheel, J. D. and Schumacher, J.}
2021
Supergranule Aggregation for Constant Heat Flux-Driven Turbulent Convection,
\textit{Phys. Rev. Research} \textbf{3}, 013231.

\bibitem[Vieweg et al. (2021a)]{Vieweg2021a}
\textsc{Vieweg, P. P., Schneide, C., Padberg-Gehle, K. and Schumacher, J.}
2021\textit{a}
Lagrangian Heat Transport in Turbulent Three-Dimensional Convection,
\textit{Phys. Rev. Fluids} \textbf{6}, L041501.

\bibitem[Vieweg et al. (2022)]{Vieweg2022b}
\textsc{Vieweg, P. P., Scheel, J. D., Stepanov, R. and Schumacher, J.}
2022
Inverse Cascades of Kinetic Energy and Thermal Variance in Three-Dimensional Horizontally Extended Turbulent Convection, 
\textit{Phys. Rev. Research} \textbf{4}, 043098.

\bibitem[Vieweg (2023)]{Vieweg2023a}
\textsc{Vieweg, P. P.}
2023
\textit{Large-scale flow structures in turbulent Rayleigh-Bénard convection: Dynamical origin, formation, and role in material transport},
PhD thesis, TU Ilmenau.

\bibitem[Vieweg et al. (2024)]{Vieweg2024}
\textsc{Vieweg, P. P., Klünker, A., Schumacher, J. and Padberg-Gehle, K.}
2024
Lagrangian Studies of Coherent Sets and Heat Transport in Constant Heat Flux-Driven Turbulent {{Rayleigh}}–{{Bénard}} Convection,
\textit{Eur. J. Mech. B-Fluids} \textbf{103}, 69--85.

\end{thebibliography}

\end{document}